\documentclass[aps,prb,reprint,superscriptaddress]{revtex4-2}
\usepackage{amssymb,amsthm,bm,bbm,booktabs,mathrsfs,mathtools}
\usepackage[dvipsnames]{xcolor}
\usepackage[bookmarks=false,colorlinks=true,allcolors=blue]{hyperref}
\usepackage{graphicx,tabularx}
\usepackage{nicematrix}

\usepackage{lipsum}

\DeclarePairedDelimiter\abs{\lvert}{\rvert}

\DeclarePairedDelimiter\ket{\vert}{\rangle}
\DeclarePairedDelimiter\bra{\langle}{\vert}
\DeclarePairedDelimiterX{\braket}[2]{\langle}{\rangle}{#1\delimsize\vert\mathopen{} #2}
\DeclarePairedDelimiterX{\matele}[3]{\langle}{\rangle}{#1\delimsize\vert\,\mathopen{}#2\,\delimsize\vert\mathopen{}#3}
\newcommand{\dyad}[1]{\ket{#1}\!\bra{#1}}

\usepackage{scalerel,relsize} 

\newcommand{\Eqref}[1]{Eq.~\!\eqref{#1}}
\newcommand{\Figref}[1]{Fig.~\!\ref{#1}}
\newcommand{\mr}[1]{\mathrm{#1}}

\begin{document}
\title{Robust Spin Qubit Coupler via Minimal Kitaev Chain}
\author{Jiaan Qi}
\affiliation{Beijing Academy of Quantum Information Sciences, Beijing 100193, China}
\author{Hongqi Xu}
\affiliation{Beijing Academy of Quantum Information Sciences, Beijing 100193, China}
\affiliation{Beijing Key Laboratory of Quantum Devices and  School of Electronics, Peking University, Beijing 100871, China}
\date{\today}

\begin{abstract}
While a minimal Kitaev chain is promised to host unprotected Majorana zero modes, its role for spin qubits is relatively underappreciated.  Following recent breakthroughs in the fine control of transport behaviors, we propose to use minimal Kitaev chain as a robust coupling module between spin qubits. Long-distance, anisotropic exchange coupling can be mediated by the Andreev bound states (ABSs) in the hybrid segment. The chemical potential of ABS gives a simple way to selectively control the coupling strength and its response to local perturbations. Moreover, this additional control degree of freedom creates a unique sweet spot, allowing both strong coupling and first-order immunity against charge noise. The protected qubit encoded on the minimal Kitaev chain at the sweet spot is shown to boast over 200 fold improvement in decoherence time. 
\end{abstract}
\maketitle 

\emph{Introduction.}---
Semiconducting spin qubits are widely conceived as compelling routes towards universal quantum computers \cite{Vaughan2023platform, Burkard2023Semiconductor,Koch2025Industrial}.  With the qubit lifetime and gate performance metrics surpassing milestones in small systems, many challenges still lie ahead in scaling up \cite{Stano2022Review,Xue2022Quantum,Noiri2022Fast,Wu2024Hamiltonian}.  In an integrated chip with multiple spin qubits, one such challenge is the control of interqubit couplings \cite{Borsoi2024Shared}. Unlike individual qubits that can be selectively addressed with at distinctive microwave frequencies \cite{Lawrie2023Simultaneous}, arbitrary control of interqubit couplings is much more demanding in terms of device design and control techniques. 

The coupling between spin qubits responsible for logic gates is typically described by the  Heisenberg exchange coupling \cite{Meunier2011Efficient,Zajac2018Resonantly}, or the anisotropic version of it under strong spin-orbit coupling (SOC) \cite{Geyer2024Anisotropic,Qi2024Spin}.  A standard method to dynamically modulate  the coupling strength involves biasing the chemical potential between neighboring quantum dots \cite{Watson2018programmable}. While this control scheme works well for small-scale 1D arrays, it tends to incur crosstalk errors as the system size increases and is cumbersome to couple multiple qubits together. This hinders advanced abilities such as implementing multi-qubit gates \cite{Qi2025Multiqubit,Gu2021Fast} or carrying out the surface code \cite{Fowler2009Highthreshold}.

Intermediatory coupling modules between qubits can be ultimate solutions to the above issues. 
This is a heated subject with growing research interests not limited to spin systems.  For superconducting qubits,  capacitively connecting pads have been verified as effective couplers \cite{Liang2023Tunablecoupling}. Bulk superconductors (SC) \cite{Spethmann2022Coupled} and flux-tunable Josephson junctions \cite{Pita-Vidal2024Strong} have also been considered for Andreev spin qubits, with a recent proposal of achieving all-to-all connectivity \cite{Pita-Vidal2025Blueprint}. For spin qubit systems, intermediatory dots or donor atoms filled with 0$\sim$2 charges  \cite{Baart2017Coherent,Rancic2017Ultracoherent, Fei2012Mediated,Deng2020Interplay,Chan2023Universal,Zhang2025Addressablea}, bulk SC leads and Josephson junctions \cite{Choi2000Spindependentb,Spethmann2024Highfidelity} have also been investigated by many as a way to mediate superexchange couplings.  
 Meanwhile, to couple spin qubits over a long distance, 
 photons in microwave cavities are also popular considerations \cite{Benito2019Optimized,Warren2019Longdistance,Harvey-Collard2022Coherent,Dijkema2025Cavitymediated}.

Hybrid quantum devices combining desirable properties of both semiconductors and superconductors (SCs) have attract great research interests in the context of quantum information \cite{Burkard2020Superconductor}. A notable example is the minimal Kitaev chain (MKC), where two normal dots are coupled by a piece of p-wave SC. MKCs are believed to host localized Majorana zero modes at specific sweet spot \cite{Leijnse2012Parity,Tsintzis2022Creatinga}, and can serve as a building block for topological quantum computers \cite{Kitaev2001Unpaired,Plugge2017Majorana}. Recently, it has also been demonstrated that SC-proximitized nanowire with strong SOC can be used in place of p-wave SC, and electronic transport through elastic cotunneling (ECT) and crossed Andreev reflection (CAR) can be accurately tunned by varying the gate voltages \cite{Liu2022Tunable,Wang2022Singlet,Bordin2023Tunable,Dvir2023Realization,Liu2024Couplingb}. 

\begin{figure}[hbp]
  \centering
  \includegraphics[width=0.95\linewidth]{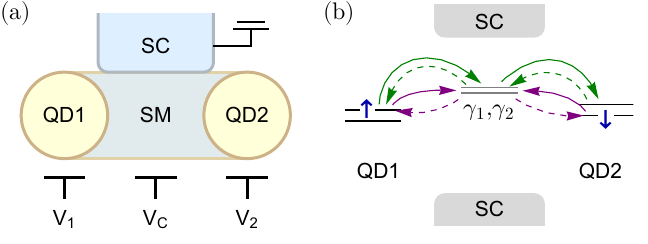}
  \caption{
(a) A minimal Kitaev chain with tunable gates for the dots and SC-proximitized nanowire. (b) The relevant spin and Andreev levels of the system under a small magnetic field. The blue and green curves (solid, then dashed arrows) illustrate the exchange paths through ECT and CAR. }
  \label{fig:dev}
\end{figure}

In this paper, we propose to use the minimal Kitaev chain as a tunable coupler between spin qubits, as shown in \Figref{fig:dev}(a). Such a coupler requires placing an extra SC lead in close contact with the semiconducting channels between normal quantum dots. SC proximity effect hybrids the middle segment and gives rise to a pair of spin-split ABS levels within the SC band gap \cite{Bauer2007Spectral,Lee2014Spinresolved}. For dot levels 
tunned to proper windows, these ABSs act as clean and gate-tunable interdot transport channels and can mediate effective exchange coupling between spins, as illustrated in \Figref{fig:dev}(b).
Unlike the usual Heisenberg exchange coupling between near-neighbors, the ABSs-mediated coupling is a non-local ``superexchange'' that does not need wave function overlap between dot states. Rather, it relies on coherent transfers between the dot states and ABSs. 
Finite spatial distributions of the ABSs allow the coupling to be transmit over a distance, limited by the SC coherence length ($\sim 1\mu$m \cite{Mayer2019Superconducting}), much longer than quantum dot radius ($\sim$50nm).  
Compared to other coupling schemes, the MKC-based couplers offer strong and gate-tunable exchange coupling over a large energy range. It has little hardware overheads and can be made robust to nuclear and charge noise. These advantages made them suitable for building scalable coupling modules for spin qubits. 

\smallskip
\emph{ABS-mediated superexchange.---} Assuming only one orbital level is tunned within the SC gap for each dot, the system can be described by the  Hamiltonian
\begin{equation}\label{eq:H_full}
	\begin{aligned}
H &= \sum_{i=1,2} H_i + H_\mr{C} + H_\mr{T}\\
&=\sum_{\mathclap{i=1,2,\mr{C}}} \ \xi_i (n_{i}-1) + \frac{\epsilon^\mr{Z}_i}{2}  (n_{i \uparrow}-n_{i \downarrow})  +\frac{U_i}{2} (n_{i} -1)^2\\
& \ \ + \gamma_\mr{C} \, d_{\mr{C} \uparrow}^+ d_{\mr{C}\downarrow}^+ + \sum_{i\neq\mr{C}} \,  (d_{\mr{C}\uparrow}^+, d_{\mr{C}\downarrow}^+)\, t_i V_i \begin{pmatrix}
  d_{i \uparrow} \\d_{i \downarrow}
 \end{pmatrix} + \mr{H.c.}.
\end{aligned}
\end{equation}
Here the hybrid segment is considered as an SC dot ``C'' with the Andreev relection rate $\gamma_\mr{C}>0$; $d_{i \sigma}$ is the field operator for the low-energy state at dot $i$ carrying spin $\sigma$, with 
$n_{i\sigma}=d_{i \sigma}^+ d_{i\sigma}$, $n_{i}=n_{i\uparrow}+n_{i\downarrow}$; the on-site energy $\xi_i$ (shifted to stabilizes a single charge at $\xi_i=0$), charging energy $U_i$ and the Zeeman splitting $\epsilon_i^\mr{Z}$ are defined for all 3 dots. Interdot tunnelings are specified by the strengths $t_i>0$ and $2\times 2$ unitary matrices $V_i$, whose non-diagonal elements can be ascribed to SOC and variations in the $g$-tensor principal axes of quantum dots. With strong hybridization, the energy scales are assumed to follow the hierarchy 
$\varepsilon^\mr{Z}_j, t_j \ll U_\mr{C},\gamma_\mr{C} \ll U_1, U_2$. Here 
$U_\mr{C}$ is small due to screening from the SC lead $\gamma_\mr{C}$ \cite{Lee2014Spinresolved}.

At vanishing $\gamma_\mr{C}$, \Eqref{eq:H_full} is reducible into invariant subspaces by the total charge number $\sum_i n_i$. With strongly coupled SC, only the parity of $\sum_i n_i$ is conserved. 
In practice, the parity may also be flipped on rare quasi-particle poisoning events \cite{Karzig2021Quasiparticle}. Here we assume that the parity lifetime is sufficiently long compared to the spin decoherence time, and focus exclusively on the even-parity sector. Experimentally, this sector can be pre-selected during system initialization by serial loading \cite{Noiri2022shuttlingbased} or Cooper-pair splitting techniques \cite{Wang2022Singlet}.

The eigenstates of uncoupled normal dots 1~\!\&~\!2 are the Fock states 
$\{\ket{0}, \ket{\uparrow}, \ket{\downarrow}, {\ket{2}\!:=\!\ket{\uparrow\downarrow}} \}$. While for dot~C, the even-parity eigenstates are mixed through local Andreev reflection by  
$\ket{g}\!:=\! u \ket{0}-v \ket{2}$ and
$\ket{e}\!:=\!v \ket{0}+u\ket{2}$ of energies 
$U_\mr{C}/2 \mp \Delta$ respectively, with 
$\Delta\!:=\!\sqrt{\xi_\mr{C}^2+\gamma_\mr{C}^2}$ and the coherence factors $(u,v)$ both depending on $\xi_\mr{C}$ \cite{Bauer2007Spectral}. 
The super-semi hybrid can be alternatively formulated in terms of ABS operators, 
\begin{equation}
	H_\mr{C} =  (\Delta'\! + \tfrac{\epsilon_\mr{C}^\mr{Z}}{2} ) \gamma_{1}^\dagger \gamma_1 +  (\Delta'\! -\tfrac{\epsilon_\mr{C}^\mr{Z}}{2} ) \gamma_{2}^\dagger \gamma_2 + U_\mr{C} \gamma_{1}^\dagger \gamma_1 \gamma_{2}^\dagger \gamma_2,
\end{equation}
with the induced gap $\Delta'\!:=\Delta-U_\mr{C}/2$. The ABS field operators $\gamma_1, \gamma_2$ are related to electronic field operators through Bogoliubov transforms. Here  
$\gamma_{1} = u d_{\mr{C} \uparrow} + v d^+_{\mr{C} \downarrow}$, 
$\gamma_{2}= u d_{\mr{C} \downarrow} - v d^+_{\mr{C} \uparrow}$ in the case of SC dot. Similar transforms can be defined for a nanowire of finite length by incorporating positional indices \cite{Liu2022Tunable}, but making such complication does not affect our main results. In the weak interaction regime $U_\mr{C}<2\Delta$, the ABSs correspond to the elementary excitations
from a common BCS ground state $\ket{g}$ of the hybrid.
The spin-resolved ABSs and doubly-excited state can be defined by 
$\ket{\uparrow}_\mr{C}:=\gamma_1^\dagger \ket{g}$, 
$\ket{\downarrow}_\mr{C}:=\gamma_2^\dagger \ket{g}$ and 
$\ket{e}:= \gamma_1^\dagger\gamma_2^\dagger \ket{g}$ accordingly.


\begin{figure}[htb]
  \centering
  \includegraphics[width=\linewidth]{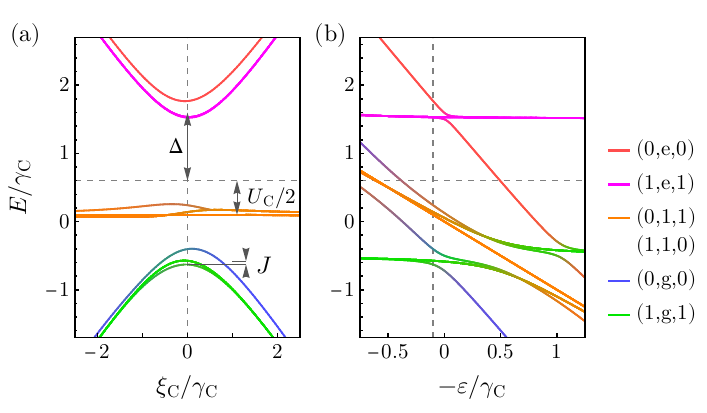}
  \caption{Energy levels of the system near $\xi_{1,2}\approx U_{1,2}/2$ at zero magnetic field, with $U_\mr{C}/\gamma_\mr{C}=1$ and $t_1=t_2=0.22\gamma_\mr{C}$. The line colors are mixed from eigenstate populations in the 6 relevant QSs (shown in the legend), with the spin subspace $(1,g,1)$ marked in green. (a) Energy levels as a function of $\xi_\mr{C}$ at $\xi_j = U_j - 0.1\gamma_\mr{C}$ for $j=1,2$, marked with the induced gap $\Delta$, charging energy $U_\mr{C}$ and the singlet-triplet energy splitting $J$. (b) Energy levels at $\xi_\mr{C}=0$ along the detuning path $\xi_{1,2}=U_{1,2}+ \varepsilon$. The vertical dashed lines in (a) and (b) share identical parameters. }
  \label{fig:elvls}
\end{figure}

Based on the smallness of $\epsilon_j^\mr{Z}$ and $t_j$, a series of quasi-degenerate subspaces (QSs) $p= (n_1,\widetilde{n}_\mr{C},n_2 )$ can be defined, with $n_1,n_2 = 0,1,2$ denoting the charge number in normal dots, and $\widetilde{n}_\mr{C}= g,1,e$ referring to the BCS ground state, single and double ABS excitations in the hybrid. 
The spin qubit pair is encoded in the $(1,g,1)$ subspace, the lowest-energy QS near the charge symmetry point $\xi_{1,2} \approx 0$. Virtual transitions to other subspaces leads to energy corrections in the spin subspace.  
Consider for example the system at $\xi_{1,2}\approx U_{1,2}$,  where influence of doubly-occupied states to the spin subspace is negligible. In \Figref{fig:elvls}, we plot 18 relevant energy levels as functions of $\xi_\mr{C}$ and $\xi_{1,2}$ at zero magnetic field, with colors mixed from the eigenstate populations in 6 different QSs. We observe from \Figref{fig:elvls}(a) that the energy bands resemble that of a typical BdG system when $U_\mr{C}\ll \Delta$, with positive energy $\ket{e}$-like states, negative energy $\ket{g}$-like states and mid-gap ABS-like states, separated by the induced gap $\Delta$. 
Within the $(1,g,1)$ manifold, a single level is split-off by a small energy $J$, accompanied by a mixing between the $(1,g,1)$ states and the $(0,g,0)$ state. 
As illustrated in \Figref{fig:elvls}(b), such a splitting is prominent near the $(1,g,1){-}(0,g,0)$ level-crossing and also at the $(1,g,1){-}(0,g,0)$ crossing, where $J$ can be neither positive or negative.

We can obtain the effective Hamiltonian for the spin subspace $H_{\mr{eff}}= H^{(0)} + \Sigma$ by applying the quasi-degenerate perturbation theory.
The zeroth order term, $H^{(0)}=\epsilon_1^\mr{Z} \sigma_1^\mr{Z}/2+ \epsilon_2^\mr{Z} \sigma_2^\mr{Z}/2$, is the projection of \Eqref{eq:H_full} onto the $(1,g,1)$ subspace responsible for defining the qubits. Virtual round-trip transitions through other subspaces leads to a perturbative series of self-energy terms $\Sigma=\Sigma^{(2)}+\Sigma^{(4)}+\cdots$. The second-order term  $\Sigma^{(2)} \propto I$ acts trivially on the spin subspace. This is expect from the intuition that separate round-trips to the middle dot cannot bring two electrons into contact. Non-trivial effects are obtained from different 4th-order exchange paths. A key finding from the perturbation is that all these paths interferes \emph{coherently} to the spin subspace, splitting-off a common state,
\begin{equation}\label{eq:exchange}
\Sigma \cong  - J \dyad{ \tilde{S} } =   J ( V_1^\dagger \bm{S}_1 V_1)  \cdot  (V_2^\dagger\bm{S}_2 V_2).
\end{equation}
Here $\ket{\tilde{S}} = V_1^\dagger \otimes V_2^\dagger \ket{S}$ 
is the effective spin-singlet rotated from the normal singlet $\ket{S}=\frac{1}{\sqrt{2}}(\ket{\uparrow,\downarrow}-\ket{\downarrow,\uparrow})$ due to 
 spin-precessed tunnelings $V_{1,2}$. The expression in \Eqref{eq:exchange}  describes an anisotropic exchange coupling between spins that can be related to the Heisenberg exchange by local frame-rotations \cite{Li2014Anisotropic,Liu2018Control}. The exchange energy $J$ can be obtained from a summation of second-order paths from $(1,g,1)$ to some subspace $p$,
\begin{equation}\label{eq:J}
  J= \sum_p J_\mr{p} = \sum_p  \frac{2{(\Gamma_p)}^2}{E_p-E_{(1,g,1)}},
\end{equation}
where the QS energy $E_{p=(n_1, g/e, n_2)}=E_1(n_1)+E_2(n_2)+\frac{1}{2} U_\mr{C}\mp \Delta$, with $E_i(0/2)=\frac{1}{2}U_i\mp \xi_i$ and  $E_i(1)=0$. The tunneling rates $\Gamma_p$ is heavily modulated by the ABS coherence factors $u$, $v$, as summarized in table \ref{tab:Jex}.
\begin{table}[hb]
  \centering
\renewcommand{\arraystretch}{1.25}
  \begin{tabularx}{0.8\linewidth}{ c >{\centering\arraybackslash}X   c}
\toprule
 $p$  & $\Gamma_p$ &  nature \\
\midrule
$(0,g,0)$ &  $ u v (j_{1-}+j_{2-}) $ & CAR \\
$(0,e,0)$ &  $ -u^2 (j_{1-}+j_{2-}) $ & CAR \\
$(0,g,2)$ &  $ u^2 j_{1-}-v^2 j_{2+}  $& ECT\\
$(0,e,2)$ &  $ u v (j_{1-}+j_{2+})$& ECT \\
$(2,g,0)$ &  $ -v^2 j_{1+}+u^2 j_{2-} $ & ECT\\
$(2,e,0)$ &  $ u v (j_{1+}+j_{2-}) $ & ECT\\
$(2,g,2)$ &  $ u v  (j_{1+} + j_{2+})$ & CAR \\
$(2,g,2)$ &  $ v^2 (j_{1+} + j_{2+})$ & CAR \\
\bottomrule
  \end{tabularx}
  \caption{List of 2nd-order tunneling rates by the target QS $p$, and the nature (ECT or CAR) of such tunneling. Here we define the shorthands that 
$j_{i,\pm}:=t_1 t_2 /(\frac{1}{2}U_i\pm \xi_i +\Delta')$.
}
  \label{tab:Jex}
\end{table}

\begin{figure}[t]
  \centering
  \includegraphics[width=\linewidth]{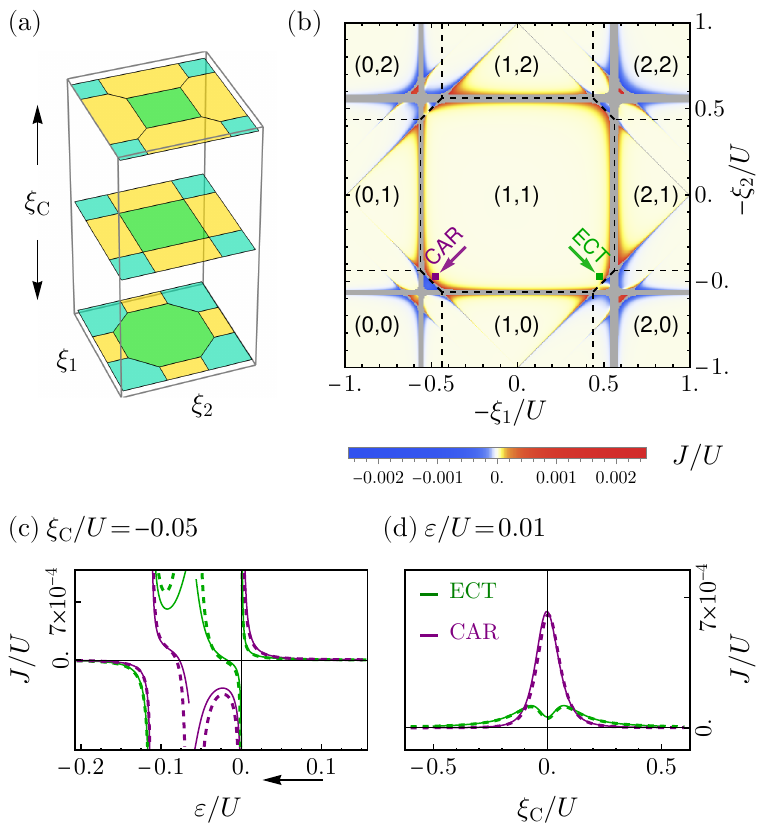}
\caption{(a) The CSD of an even-parity MKC controlled by $\xi_\mr{C}$, with the $(1,1)$ region in green and the $(0,0)$, $(0,2)$, $(2,0)$, $(2,2)$ regions in cyan. 
(b) The exchange energy $J$ overlaid on a CSD slice ($\xi_\mr{C}=-0.05 U$) for a typical MKC. The dashed lines represent the CSD boundaries and gray shades are level-crossing regions where the spin subspace cannot be defined [with $(1,g,1)$ population $\ge 0.9$]. The green and purple arrows indicate detuning paths to the $(2,0)$ and $(0,0)$ regions, where the exchange coupling is contributed by ECT and CAR processes. 
(c) $J$ along the ECT/CAR detuning path $(\xi_1,\xi_2)=(\mp U/2 \pm \varepsilon, U/2-\varepsilon)$, with $\varepsilon \to 0^+$ corresponding to the green/purple arrow in (b). The paths are extended to also allow $\varepsilon<0$ values beyond the $(1,1)$ region. 
(d) Modulation of $J$ by $\xi_\mr{C}$ at the fixed points $\varepsilon=0.01U$ on the ECT and CAR path, as marked with the green and purple squares in (b). The solid and dashed lines in (c) and (d) are from plotted from numerical (energy splitting) and theoretical [\Eqref{eq:JECTJCAR}] values respectively.
}
 \label{fig:Jnum}
\end{figure}

\emph{Control schemes.---}
The operational state of a double-dot system can be visualized with a point in the charge stability diagram (CSD), which specifies the ground-state charge number $(n_1, n_2)$ over the dot potentials $(\xi_1,\xi_2)$. For our system with gate-tunable ABSs, $\xi_\mr{C}$ serves as an extra dimension for the CSD, as illustrated in \Figref{fig:Jnum}(a), that controls the region boundaries as well as exchange energy $J$. 
In \Figref{fig:Jnum}(b), we visualize the distribution of $J$ on the 2D CSD slice at $\xi_\mr{C}=-0.05 U$ for a MKC parameterized by 
$U_{1,2}=U$, $\gamma_\mr{C}=0.1U$, $U_\mr{C}=0.05U$, $t_{1,2}=0.01 U$ and $V_{1,2}=R(\pm 0.125\pi)$, for the 2D rotation matrix $R(\theta)$. The $J$ values are numerically extracted from the singlet-triplet energy splitting within the $(1,g,1)$ subspace and plotted under a non-uniform colorbar to highlight key features. Within the $(1,1)$ region of the CSD, $J>0$ is tiny everywhere except near the boundaries towards $(0,0),(0,2),(2,0)$ and $(2,2)$. In regions other than $(1,1)$, the spin subspace is no longer lowest in energy and $J$ can be either positive or negative. There are also certain points (gray-shaded areas) with undefinable $J$. These corresponds to level-crossing regions where a spin subspace cannot be isolated with sufficient fidelity due to heavy mixing between (1,g,1) and other QSs.
Spin qubits should be operated away from these level-crossing regions to avoid leakage errors.

Effective control of interquit couplings requires switching between a $J\approx 0$ point and a large-$J$ point in the control space $(\xi_1,\xi_2,\xi_\mr{C})$. 
Two types of control schemes can be naturally defined. 
In the first detuning-based scheme, one fix $\xi_\mr{C}$ and move the dot potentials along an engineered path in the CSD. For example, following the arrows in  \Figref{fig:Jnum}(b), $J$ quickly picks up magnitudes when approaching level-crossing boundaries. This can be seen by taking $\varepsilon\to 0^+$ from the positive direction in \Figref{fig:Jnum}(c), where $\xi_\mr{C}=-0.05 U$ and $J$ is varied along the ``ECT'' path 
$(\xi_1,\xi_2)=(- \frac{1}{2} U + \varepsilon, \frac{1}{2}U-\varepsilon)$ in the $(2,0)$ direction and the ``CAR'' path 
$(\xi_1,\xi_2)=(\frac{1}{2} U- \varepsilon, \frac{1}{2} U-\varepsilon)$ in the $(0,0)$ direction. This scheme is comparable to detuning-based controls in regular double-dots, with the additional CAR paths made possible by the ABSs. 
Alternatively, one can adopt the ABS-based control scheme, that modulates $J$ by tuning $\xi_\mr{C}$ with the dot potentials $(\xi_1,\xi_2)$ fixed.
This is demonstrated for the two points indicated with square markers in \Figref{fig:Jnum}(b), with $\varepsilon=0.01 U$ along the ECT \& CAR path. The responses of $J$ against $\xi_\mr{C}$ in the two cases are plotted in \Figref{fig:Jnum}(d).

At small detunings, the exchange coupling is primarily contributed by CAR transitions from $(1,g,1)$ to the ${(0,g,0)}$ and ${(0,e,0)}$ QS in the $(0,0)$ sector and ECT transitions to ${(2,g,0)}$ and ${(2,e,0)}$ in the $(2,0)$ sector. This allows us to simplify \Eqref{eq:J} into 
\begin{equation}\label{eq:JECTJCAR}
\begin{aligned}
		J_\mr{CAR} &\simeq
\frac{t_1^2\, t_2^2}{ (\varepsilon\, {+}\, \Delta')^2 }
\left[\, \frac{1}{\varepsilon} \Bigl(\frac{\gamma_\mr{C}}{\Delta}\Bigr)^2 \! \!+ \frac{1}{\varepsilon+\Delta} \Bigl(1+\frac{\xi_\mr{C}}{\Delta}\Bigr)^2\right],\\[1ex]
		J_\mr{ECT} &\simeq \frac{t_1^2\, t_2^2}{ (\varepsilon\, {+}\, \Delta')^2 }
\left[\, \frac{1}{\varepsilon} \Bigl(\frac{\xi_\mr{C}}{\Delta}\Bigr)^2 \! \!+ \frac{1}{\varepsilon+\Delta} \Bigl(\frac{\gamma_\mr{C}}{\Delta}\Bigr)^2\right].
\end{aligned}
\end{equation}
In \Figref{fig:Jnum}(c-d),
The $J$ values obtained from energy splittings (solid lines) agree closely with the predictions from the above (dashed lines),
 verifying the validity of our analysis. Furthermore, in the $\varepsilon\ll\Delta$ regime, the above expressions are dominated by the first terms in the square brackets. As a result, the CAR-induced coupling is maximized around $\xi_\mr{C}=0$, where the ECT-induced coupling sees a dip [see \Figref{fig:Jnum}(d)]. When $\abs{\gamma_\mr{C}}\gg \gamma_\mr{C}$, both ECT and CAR are suppressed by the ABS energy barrier $\Delta$, thus effectively switching off the couplings.

\emph{Sweet spot and protected qubit.---} 
Apart from providing novel control leverages, another reason making MKC-based couplers worthy of 
 consideration is the presence of protected sweet spots, with both large coupling and immunity against charge noise. Hence a primary source of error for spin qubits can be mitigated.

\begin{figure}[htbp]
  \centering
  \includegraphics[width=\linewidth]{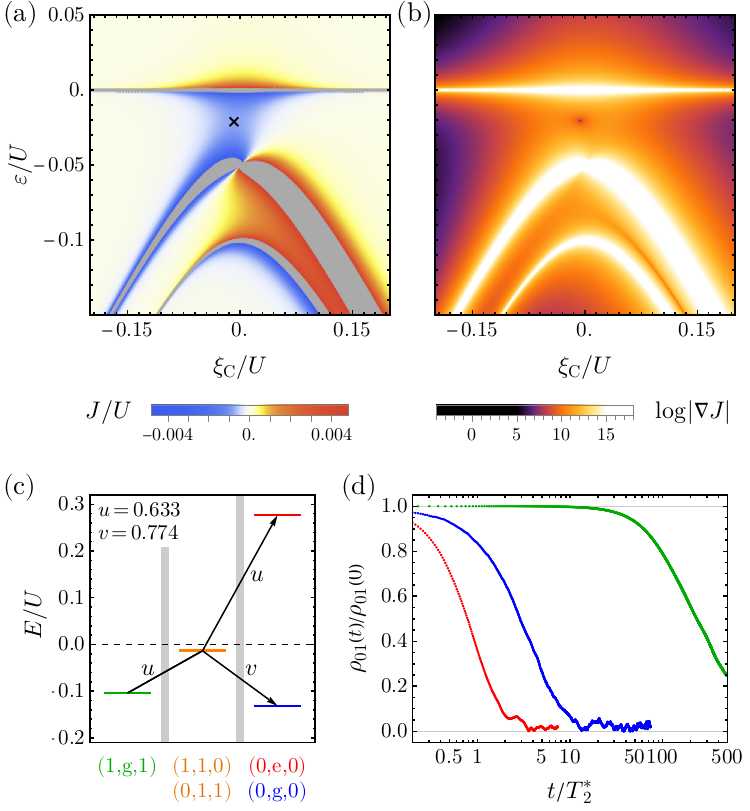}
  \caption{(a) $J$ for the same MKC as in \Figref{fig:Jnum}(b-d), plotted against $\xi_\mr{C}$ and $\varepsilon$ along the CAR path 
 $(\xi_1,\xi_2)=\frac{U}{2}-(\varepsilon,\varepsilon)$. The 4 colored regions separated by level-crossings (gray areas) correspond to different QS energy orderings. At the black cross, $\partial J/\partial \xi_\mr{C}=\partial J/\partial \varepsilon =0$, with $J<0$  relatively large magnitude. (b) Norm of the 3D gradient $\nabla J = (\partial_{\xi_1} J, \partial_{\xi_2} J, \partial_{\xi_\mr{C}} J)$ in log scale. A sweet spot with vanishing gradient emerges at half-way between level-crossings, serving as a ``in-gap'' level protected against charge noise.
(c) The QS energy levels and ABS coherence factors at the sweet spot. In such configuration, the $(1,g,1)$ subspace is separated from the ground state $(0,g,0)$ by second-order tunneling across the ABS energy barriers. (d) Benchmarking the traversal coherence factor $\rho_{01}$ of qubits prepared in a noisy environment with both magnetic and charge noises. The red, blue, and green lines represent performances for the base-line spin qubit, $S{-}T_0$ qubit and the Kitaev-spin qubit [\Eqref{eq:qubit}] operating in the sweet spot respectively. The environment noises are modelled with $1/f$ stochastic Gaussian processes with noise strengths estimated from  related experiments.
}
  \label{fig:sweet}
\end{figure}

To understand why current spin qubit architectures are sensitive to charge noise, consider the exchange coupling $J\simeq 2t^2/(U+\xi_1-\xi_2)$ near the $(1,1){-}(2,0)$ boundary. $J$ can be ``turned-on'' with the detuning $\xi_1-\xi_2 \to- U$, similar to the $\epsilon \to 0^+$ limit in \Figref{fig:Jnum}(c). Large $J$ is accompanied by a large gradient $\partial J/\partial \xi_i$ near the boundary. As a result, small environment-induced fluctuations in  $\xi_i$ are reflected as large variations in $J$. This uncertainty creates incoherent errors, which are further exacerbated by population leakage errors into the $(2,0)$ state. 

To overcome the above problem, one can operate the MKC coupler at a special sweet spot in the control space, as demonstrated by the black cross in \Figref{fig:sweet}(a), where we plot $J$ along the CAR path $(\xi_1,\xi_2)=\frac{U}{2}-(\varepsilon,\varepsilon)$
for the same MKC considered in \Figref{fig:Jnum}. The black cross is the unique intersect of $\partial J/\partial \xi_\mr{C}=0$ and $\partial J/\partial \varepsilon =0$ for $J=J_\mr{ECT}$ in \Eqref{eq:JECTJCAR}. 
We plot in \Figref{fig:sweet}(b) the vector norm of the gradient 
$\nabla J:= (\partial J/\partial \xi_1, \partial J/\partial \xi_2, \partial J/\partial \xi_\mr{C})$ in the 3D control space for the same setup. A sweet spot is numerically found at $(\xi_\mr{C},\varepsilon)\simeq(-0.0068U,-0.020U)$, around half-way between two level-crossings. On the sweet spot, $J/U \simeq -8.5\times 10^{-4}$ and 
$\abs{\nabla J}\simeq 0$ (with the numerical precision of $10^{-20}$). Comparing to $J/U=6.1\times 10^{-4}$ and 
$\abs{\nabla J}\simeq 7.2\times 10^{-4}$ at the CAR boundary $\varepsilon=0.01 U$ and 
$\xi_\mr{C}=0$, the sweet spot has stronger coupling and first-order protection against charge noise.

A potential downside for operating at the sweet spot is that the spin subspace is no longer the ground QS. As illustrated in \Figref{fig:sweet}(c), $(1,g,1)$ is now the second-lowest-energy QS of the system, separated with the ground QS $(0,g,0)$ by second-order tunneling barriers provided by the ABSs. Moreover, due to energy conservation, the CAR current $(1,g,1)\to (0,g,0)$ is only significant
at the $\varepsilon=0$ boundary, with a finite broadening determined by the transition rate $\Gamma_{(0,g,0)}$ \cite{Liu2022Tunable}. 
Since the sweet spot is located at a safe distance 
$\abs{\varepsilon} \gg \Gamma_{(0,g,0)}$ away from level crossings, leakage into the ground state requires participation of high-energy bosonic environment modes which are exponentially suppressed at low temperatures \cite{Schlosshauer2007}. 

The sweet spot for a MKC-based coupler satisfies 3 desirable properties: (1) $\abs{J}$ is sufficiently large to allow strong interquit coupling; (2) The gradient $\nabla J$ is vanishingly small to avoid charge errors; (3) The system is operated away from level-crossing boundaries to avoid leakage errors. Curiously, a point satisfying the above 3 conditions simultaneously can only be found in the CAR regime of a MKC. Spots with vanishing $\nabla J$, for example, also exist in empty-dot couplers \cite{Baart2017Coherent,Rancic2017Ultracoherent,Chan2023Universal}. This correspond to the $\sum_i n_i=2$ and $\gamma_\mr{C}=0$ case in our model with only ECT-induced exchange. But the $J$'s are too small to be useful in these cases.

Consider a two-level system defined through 
\begin{equation}\label{eq:qubit}
\begin{aligned}
		\ket{0}&:= V_1^\dagger \otimes V_2^\dagger (\ket{\uparrow, g,\downarrow}+\ket{\downarrow,g,\uparrow}),\\
	\ket{1}&:= V_1^\dagger \otimes V_2^\dagger (\ket{\uparrow,g, \downarrow}-\ket{\downarrow,g,\uparrow})+ \epsilon \ket{0,g,0},
\end{aligned}
\end{equation}
where $\epsilon$ represents a small population mixture due to a finite CAR rate. In theory, when operating at the sweet spot, this ``Kitaev-spin'' qubit provides a decoherence-free-subspace immune to most error sources common to spin qubits. First, magnetic noise due to hyperfine interactions is suppressed, since the qubit is encoded in a net-zero-spin subspace similar to the $S{-}T_0$ spin qubits. Second, exposure to charge noise is avoided at the sweet spot due to the vanishing $\nabla J$. Third, leakage errors into non-computational subspace is prevented by operating away from level crossing boundaries.
In \Figref{fig:sweet}(d), we benchmark the traversal coherence time $T_2^*$  for 3 different qubit encodings: regular (Loss-Divincenzo) spin qubit, $S{-}T_0$ qubit and the Kitaev-spin qubit defined above, in a simulated setup with both magnetic and  charge noise. We model these noises by stochastic fluctuations in $\varepsilon_{i}^\mr{Z}$ and $\xi_i$ following a Gaussian process with $1/f$ spectrum, with strength estimated from related experiments. While the  $S{-}T_0$ encoding offer protection from the magnetic noise, it suffers from charge noise and leakage errors. In comparison, the Kitaev-spin encoding at the sweet spot condition enjoy over 200 fold improvement in the $T_2^*$ time compared to the base-line $S{-}T_0$ qubit. 

\emph{Robust qubit coupling modules.---}
MKCs combined with ABS-based control are suitable for creating selective coupling modules in large-scale chips. Consider a 2D array of $n^2$ quantum dots interconnected with a grid of $\sim2n^2$ MKC-based couplers. The coupling status between qubits is controllable by the chemical potentials of ABSs in the couplers, which can be selectively controlled using a classical circuit of $\sim 2 \sqrt{2} n$ pins. Meanwhile, crosstalk errors common to normal dot arrays are diminished in this arrangement.
First, the interdot charging energy terms $U_{i j} n_i n_j$ in the Hamiltonian are drastically screened-off by the SC leads. This isolated each qubit from charge disturbances $\delta n_j$ in neighboring sites. Second, with ABS-based control, the dot potentials remain static throughout the entire process, hence reducing any undesirable wave function polarization introduced by tilting the dot potentials. Last but not least, the MKC can be tunned to a protected sweet spot robust against number fluctuations in the local voltages, allowing quantum gates to be performed with high fidelity.

\emph{Summary.---} We investigate the possibilities brought by incorporating the minimal Kitaev chain into spin qubit systems. It is found that they can serve as efficient couplers for the exchange couplings with the gate-tunable ABS chemical potentials. We derive the exchange energy using 4th order perturbation theory and discover the existence of charge-noise insensitive sweet spot at the CAR regime. The various merits of MKCs making them ideal for robust coupling modules between spins.

\begin{acknowledgments}
J. Q. acknowledges support from the National Natural Science Foundation of China with Grant No.~12404562. 
H. Q. X. acknowledges support from  the National Natural Science Foundation of China (Grant Nos.~92165208 and 11874071).
\end{acknowledgments}

\bibliographystyle{apsrev4-2}
\bibliography{coupler.bbl}

\end{document}